# LA-UR-

Title:

Author(s):

Submitted to:

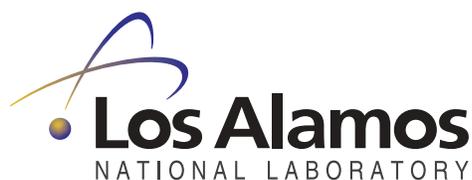

# Absence of magnetic moments in plutonium

J. C. Lashley[1], A. Lawson[1], R. J. McQueeney[2], and G. H. Lander[3]

[1] Los Alamos National Laboratory, Los Alamos, New Mexico 87545, USA
[2] Ames Laboratory and Department of Physics and Astronomy, Iowa State University, Ames, Iowa 50011, USA
[3] European Commission, JRC, Institute for Transuranium Elements, Postfach 2340, Karlsruhe, Germany

## Abstract

Many theories published in the last decade propose that either ordered or disordered local moments are present in elemental plutonium at low temperatures. We present new experimental data and review previous experimental results. None of the experiments provide any evidence for ordered or disordered magnetic moments (either static or dynamic) in plutonium at low temperatures, in either the α- or δ-phases. The experiments presented and discussed are magnetic susceptibility, electrical resistivity, NMR, specific heat, and both elastic and inelastic neutron scattering. Many recent calculations correctly predict experimentally observed atomic volumes, including that of δ-Pu. These calculations achieve observed densities by the localization of electrons, which then give rise to magnetic moments. However, localized magnetic moments have never been observed experimentally in Pu. A theory is needed that is in agreement with all the experimental observations. Two theories are discussed that might provide understanding of the ensemble of unusual properties of Pu, including the absence of experimental evidence for localized magnetic moments; an issue that has persisted for over 50 years.

## I INTRODUCTION

It has been known for many years that plutonium lies in the periodic table at a position where it is intermediate between itinerant- and localized-electron behavior.[1] The elemental volumes of the 5*f* elements are shown in comparison to those of the elements in the 3*d* and 4*f* series in **Fig. 1**. The behavior of the early actinides (Th to Np) follows closely the *contraction* with increasing electron count that is systematically followed in all the *d* transition-metal series. At the beginning of the series each additional electron contributes to the cohesive energy of the solid, resulting in a *decrease* of volume until the shell is approximately half full. This characteristic of the early actinides, together with the absence of magnetic order, has been taken as a *prima fascia* case that the 5*f* electrons of these early actinide elements are *itinerant*. On the other hand, for the heavier actinide elements, there is an abrupt (at δ-Pu and Am) jump in the volume and very little change as the electron count is further increased. In comparison with the 4*f* elements, together with the presence of ordered magnetism in Cm and the elements beyond (those that have been examined), this change in trend has been taken as evidence of *localized* behavior of the 5*f* electrons. If we accept this hypothesis, then it focuses a major interest on plutonium. Note that the volume change between α-Pu and Am is almost 50%, a staggering change in volume between two neighboring elements in the periodic table considering that the only change is to add one electron in the 5*f* shell. (Unlike the lanthanide elements Eu and Yb, which are both divalent in the normally trivalent lanthanide series, there is no indication of a straightforward valence change between Pu and Am)

Plutonium, however, not only has the α-Pu phase that clearly falls on the "itinerant-like" volume line of Fig. 1, but it also exhibits the δ-Pu phase with a volume expansion of ~25% as compared to the α-phase. It is further known that by adding a small amount (a few per cent) of Ga or Al to the α-Pu phase, the simple fcc δ-Pu phase can be stabilized and thus studied at room temperature (and below). The extraordinary properties of plutonium metal are well illustrated by the thermal expansion shown in **Fig. 2**. This shows dramatically the large increase (>20% in volume) between the α and δ phases.

Experimental work on plutonium metal demands special facilities because of the toxic and radioactive nature of the element. This has confined the experimental studies to a small number of institutions around the world. In addition, not all the experiments have been published in readily accessible journals, so the overall situation with respect to reliable experimental evidence of transitions below room temperature is at best vague and, at worst, confusing.

No such constraints lie on the theory side. As we summarize in Sec. II, the theory community have been especially active in the last ~ 20 years, with increasing contributions in the last 5 years, and a large number appear to believe that in order to explain the volume differences between α- and δ-Pu, there simply *must* be ordered magnetism in δ-Pu. Indeed, various antiferromagnetic (AF) structures have been proposed comparing the ground-state energies. The controversial aspect of this discussion is that the experimental evidence for magnetic moments on the Pu atoms is almost non-existent.

We show in Sec. III the most recent experiments on the low-temperature properties of δ-Pu. We maintain that there is *no evidence whatsoever* that magnetism (either ordered or disordered) exists in δ-Pu. We draw this conclusion from a series of past experiments as well as the latest studies. A similar conclusion is reached about the α-Pu phase.

Section IV draws some conclusions and highlights the challenge; *why is there no magnetism appearing in this phase?*

## II  THEORY OF THE GROUND STATE OF PLUTONIUM

The theory of Pu has been addressed in considerable detail in a number of articles in Los Alamos Science.[1] Relativistic, but not self-consistent, calculations were already performed in the 1970s and a review of all work up to ~1983 can be found in Brooks, Johansson and Skriver.[2] Very early in the development of the theory it was realized that by allowing the 5*f* electrons in Pu to spin polarize in the calculation the volume would be increased. The specific example of this process in Am is discussed in Ref. 2.

Schadler *et al.*[3] in 1986 discussed spin polarization in the ground state of Pu metal, however, they did not consider the effects of orbital polarization. Soon after the above paper, some of the aspects of the thermal expansion were treated by Söderlind *et al.*[4] and relatively good agreement was obtained for the δ-phase when the relativistic spin-orbit interaction was included in the calculations. The first specific mention of ordered magnetism in the δ-phase that we could find is in the paper by Solovyev *et al.*[5] in 1991. These authors obtained spin and orbital moments of 3.5 $\mu_B$ and –2.0 $\mu_B$, respectively, giving a net moment of 1.5 $\mu_B$. Another interesting point about these calculations is that the orbital moment is less than the spin moment, $|\mu_L| < |\mu_S|$. This is a feature we shall find for *all* the calculations made by band theory. In the normal application of Hund's rules to the light (less than half filled) *f*-states we should find the opposite situation, $|\mu_L| > |\mu_S|$. This aspect is further discussed by Hjelm *et al.*[6] in an interesting paper about the induced magnetic form factor in the light actinides. Unlike the case of uranium, where the spin and orbital moments couple ferromagnetically when a large magnetic field is applied (in agreement with experimental results taken years before) the larger spin-orbit interaction in Pu ensures that $\mu_L$ and $\mu_S$ are antiparallel *and* that $|\mu_L| > |\mu_S|$.

The fundamental reason why the theories arrive at magnetic order is that localization of the 5*f* states expands the atomic volume and thus reproduces the expansion from the itinerant "line" (see Fig. 1) to the localized "line". The *localization* of the 5*f* states immediately meets the volume criterion of the δ-phase, and thus arrives at a $f^{5/2}$ shell that is nearly full with five 5*f* electrons. If this is allowed to spin polarize a large magnetic moment is obtained.

One of the first studies to show this explicitly was that of Eriksson *et al.*[7] in 1992 who used calculations of the electronic structure of atoms on the *surface* of α-Pu. Because atoms at the surface can relax to the vacuum, they can change their volume; hence a prediction of δ-like surface states with a magnetic moment. Not surprisingly, large moments were found by the theory; they were dominated by the spin moments of some 4 – 5 $\mu_B$. Eriksson *et al.* predicted surface magnetism. In 1995 Antropov *et al.*[8] made comparisons between α-Mn and α-Pu and suggested that the bulk α-phase *also* had large magnetic moments, again dominated by the spin terms, of between 1 – 2 $\mu_B$.

Chronologically, we note the paper of Nordström & Singh[9] in 1996, where they predicted an unusual "noncollinear intra-atomic magnetism" in Pu. We shall return to this paper later.

Although the GGA method was first applied to the Pu case in 1994 by Söderlind *et al.*[10], it was not until the LDA+*U* papers in 2000 by Savrasov and Kotliar[11] and Bouchet *et al.*[12] that the large magnitude of the moments predicted in the δ-phase became clear. Bouchet *et al.*[12] pointed out the contradiction of the large moments emerging from these calculations and the paucity of experimental data to corroborate them.. They emphasized the compensation of the spin and orbital moments and the dominance in the calculations of the spin moment as long as the 5*f* states are treated as itinerant. However, as the localization occurs, there is increasing *orbital polarization* so that finally one must, with five 5*f* localized states, tend to the Russell-Saunders coupling result. If intermediate coupling is considered, this gives $g = 0.414$ (the Landé factor) and $\mu_L = + 3.97\ \mu_B$ and $\mu_S = - 2.93\ \mu_B$, resulting in $\mu_{total} \sim 1\ \mu_B$.

At the same time, Eriksson et al[13] advanced a theory for bulk δ-Pu in which only *some* of the 5*f* spectral weight was localized. Similar ideas were also advanced by Cooper et al.[14]. In neither of these theories was ordered magnetism predicted.

The cancellation aspect of the spin and orbital moments was discussed by Savrasov & Kotliar,[11] but a significant contribution came a year later[15] when they reported calculations with the dynamical mean-field theory (DMFT) and postulated that there was *no* static ordering in δ-Pu because the fluctuation time of the ordered moments was too short. We shall return later to discuss the possible implications of this.

Detailed ground-state magnetic configurations of plutonium were published by Postnikov and Antropov[16] in 2000 with moments between 0.25 and 5 $\mu_B$, and this was followed by Wang and Sun,[17] who proposed AF solutions, although they do not give the value of the ordered moments.

From 2001 onwards, Söderlind and collaborators have published a series of papers predicting magnetism in both α- and δ-plutonium. The first paper[18] invoked the cancellation of $\mu_L$ and $\mu_S$ and arrived at ~ 3 $\mu_B$ for both quantities. The volume calculated for δ-Pu was very close to the experimental value, as were the results of Ref. 11. A more detailed work following the same line was published a year later[19] in which the authors find a resultant moment of ~1.5 $\mu_B$, and claimed that the δ-phase had *disordered* moments of approximately this value. In 2003 the same authors went on to consider α-Pu with the same theory[20] and concluded that it too was an antiferromagnet. Because α-Pu has many different sites in the unit cell, the moments predicted varied between 0.5 and ~ 3 $\mu_B$. About the same time Kutepov and Kutepova,[21] obtained similar results, finding AF ground states for both α- and δ-Pu, with values of the moments up to ~ 1.5 $\mu_B$ for the α-Pu and almost 2.5$\mu_B$ for the δ-phase. In all cases, as in the previous calculations, the results are dominated by the *spin* moments, i.e. $|\mu_L| < |\mu_S|$. Landa and Söderlind[21] have recently invoked the magnetic entropy as the origin of the stabilization of δ-Pu when small small amounts of *p* elements such as Al and Ga are added.

Niklasson et al.[23] have also discussed the modeling of actinides with disordered local moments (DLM). This paper summarizes very well the present theoretical situation with respect to the *volume* and *magnetism* of the actinide metals, especially plutonium. They point out the vast improvement in both the volumes and the bulk moduli when disordered magnetism is allowed for δ-Pu. They find a moment of ~ 4.5 $\mu_B$ at the Pu site, but they expect this to be reduced to ~ 2 $\mu_B$ if correct spin-orbit coupling and orbital polarization is included. They conclude, however, by making the remark: "... the intention of the DLM picture is to model some of the main characteristics of the energetics of the actinides, and it does not necessarily describe the magnetic properties correctly."

A new theory paper[24] emphasizing the importance of magnetism in all phases of Pu appeared in 2004, as this work was submitted. The authors dismiss the disagreement between experiment and their theory with the sentence: "This fact strongly suggests that magnetism plays a role in Pu although screening or other effects may obscure its existence experimentally."

The above summary of the current state of theory on plutonium is by no means exhaustive. Many other papers have been written on theories[1] describing the complex physical properties of plutonium; our aim has been simply to summarize those (numerous) papers predicting a magnetic ground state.

The clear consensus of a large body of theoretical work on this subject is that to understand the large volume expansion between α- and δ phases of plutonium a *localization* of the 5*f* states is required, and this leads inexorably to the prediction of magnetic ordering for the δ-phase. In some cases, magnetic ordering is also found to occur in calculations on the α-phase. In both cases the resulting moments are predicted to be large, and even if a partial cancellation occurs for the spin and orbital parts, the resulting (static) magnetic moments are of the order of 1 to 2 $\mu_B$, depending on the details of the calculations. We now examine the published experimental evidence for magnetism, either disordered or ordered in both α- and δ-Pu.

## III     EXPERIMENTS ON PLUTONIUM

In contrast to the large number of theory papers mentioned above (and there are an equal number not discussing magnetism that we have not cited) the experimental situation with respect to Pu is sparse. Most experiments examining the properties below room temperature were performed in the 1960s. In many cases the samples were not of the highest purity, and the results were often published in conference proceedings or in journals read by metallurgists rather than physicists. This has led to a far from evident literature, and indeed it is often hard to find copies of some of these conference proceedings 40 years later. Fortunately, some good reviews have been written. One of the best is the article by Lee & Waldren[25] in (1972) where many of the properties of the metals are summarized as presented at the "Plutonium and other Actinides" Conference held in Santa Fe, New Mexico in October 1970.[26] These early works noted anomalies in α-U (at 43 K) and in α-Pu (at 60 K) but did not generally invoke magnetism, or the localization of the 5*f* electrons in δ-Pu, for example.

Seven important experimental papers were published on possible magnetism in Pu in the period 1960 – 1972.

(1) Sandenaw[27] and coworkers published a study of the specific heat of α-Pu below 420 K in 1960, and a study[28] of stabilized δ-Pu with 8 % Al about the same time. Results in the α-phase were characterized by several peaks, whose origin below 100 K was attributed to a level splitting of the 5*f*-electrons (analogous to the Stark splitting of 4*f* -electrons observed in lanthanides). The peak centered at 123 K was attributed to the presence of spin disorder, that is to say a transformation out of an antiferromagnetic state. However, antiferromagnetic ordering in the specific heat of α-Pu was later ruled out after the Plutonium 1965 meeting in an analysis published by Taylor and Linford[29] in 1967.

(2) Similar results were reported for the specific heat of the δ-phase with four peaks loacated at 31, 45, 62, and 190 K. So similar in fact that Sandenaw suggested that a possible explanation was that the specific-heat behavior was not a property of the crystal structure but of the Pu atoms themselves. It was later pointed out by Taylor et al.[30] that the source of the peaks in Sandenaw's measurements were artifacts of the technique because the measurements were made in exchange gas, some of which adsorbed onto the sample during the measurement.

(3) Brodsky[31] published a study of the magnetoresistivity in α- and β-Pu, and measured a negative effect. Although a negative magnetoresistance is generally associated with antiferromagnetism it can also be a result of weak localization in a low-dimensional system. He postulated $T_N \sim 27$ K, even though there was no discontinuity in the resistivity curve at this temperature.

(4) In 1970 Fradin & Brodsky[32] published an account of NMR experiments looking at the $^{27}$Al nucleus in a δ-Pu sample stabilized with 4% Al. The results showed no sign of any reduction of the local symmetry at any temperature, thus arguing against any magnetic ordering.

(5) Fournier[33] published a paper on the susceptibility, and surveyed some of the earlier results, on α-Pu and postulated "almost" magnetic behavior in this material at low temperature.

(6) Blaise & Fournier,[34] based on an analysis of the susceptibility of α-Pu, postulated that at 60 K the 5$f$ electrons become localized and thus local moments exist, but they are disordered down to 4.6 K.

(7) In 1972 Arko et al.[35] on the basis of resistivity measurements for both α- and δ-Pu, and an extensive review of all other measurements up to that time, concluded that there was *no evidence for localized moments or magnetic order*. On the basis of a $T^2$ dependence of the resistivity near $T = 0$, they suggested a model involving spin fluctuations for elemental Pu, as well as for a number of other alloys and compounds. The theory of electrical resistivity caused by spin fluctuations in the actinides was developed by Kaiser and Doniach[36] in 1970.

We shall now examine in detail some of the individual physical properties and possible evidence for magnetism, either disordered (static or fluctuating) or true long-range ordered.

### A     Magnetic Susceptibility

**Figure 3** shows the molar magnetic susceptibilities of Mn and Pu metals plotted versus temperature, scaled by melting point, $T_m(Mn) = 1519$ K, $T_m(Pu) = 913$ K. Manganese is chosen for comparison to plutonium because the two metals have similar values of susceptibility, as first pointed out by Sandenaw,[37] and both metals undergo a succession of phase transformations through progressively less complex crystal structures as the temperature is raised. Starting from 300 K in the monoclinic α-structure, Pu transforms to β (also monoclinic), γ (orthorhombic), δ (FCC), ε' (body centered tetragonal, BCT), and ε (BCC) structures.  The sequence for Mn is α (complex cubic), β (also complex cubic), γ (FCC) and δ (BCC).   Mn orders antiferromagnetically at 95 K ($T/T_m = 0.063$) in a complex tetragonal structure that is a slight distortion of the α-phase.[38,39]

The Mn data are taken from Kohlhaus and Weiss,[40] as reviewed by Wijn.[41] For elemental Pu above 300K, the data are those of Comstock, published posthumously in the article by Sandenaw,[37] and in further detail by Olsen *et al*.[37] These data were obtained by the accurate Gouy method on a large sample of $^{239}$Pu with a stated purity of 99.9%. (Unfortunately, it is not known whether this specification expresses weight or atomic percent; this is significant because 0.15 % Ni by weight is sufficient to suppress the appearance of the δ' phase.[42]) Low-temperature data for α-phase and for stabilized δ-phase (Pu – 6 at. % Ga) are from the paper by Méot-Raymond and Fournier[43] as are high-temperature data for the Pu-Ga alloy. In the latter case, a correction factor of 1.139 has been applied to align the low- and high-temperature data; this procedure seems to be justified by experimental concerns mentioned by Méot-Raymond and Fournier[43] concerning their high-temperature measurements. With this correction, the agreement between susceptibility values for unalloyed δ and stabilized δ is good.

In general it is important to note that the susceptibility is large, especially for an element. Various theories have treated this result in terms of narrow bands cutting the Fermi level. Such an interpretation is consistent with the unusual behavior of the resistivity of the early actinide metals.[25] There is no sign of any anomaly at low temperature, as might be associated with magnetic ordering. In the case of Mn, the susceptibility increases on cooling after the α-phase is formed at $T/T_m = 0.63$ and there is a clear anomaly at $T_N = T/T_m = 0.063$. The surprising aspect of the results for Pu is that the change in the molar susceptibility between the α- (or stabilized δ) phase at low temperature and the ε-phase at high temperature is *small*. This measurement gives little credence to the idea that local moments are developing below $T/T_m \sim 0.4$, whether in the α- or stabilized δ-phase. In particular, the attempt by Méot-Raymond and Fournier[43] to analyze their data in terms of a large $\chi_o$ (*T*-independent term) and then a *T*-dependent contribution reflecting the local moments, can be seen to be fraught with considerable danger. The $\chi_o$ term comprises over 85% of the measured susceptibility, casting doubt on the resulting "effective moments" deduced from the remaining susceptibility for a number of stabilized phases. Unfortunately, the value of $\mu_{eff} \sim 1.2\ \mu_B$ deduced from this analysis of the susceptibility in stabilized (6 at % Ga) δ-Pu has been invoked by a number of theorists to justify their predicted local moments in plutonium. Fig. 3 makes it clear that such a conclusion is far from evident when the whole molar susceptibility curve versus temperature is considered.

Measured magnetic susceptibilities of Pu in its various phases are characteristic of metals with relatively strong paramagnetism caused by electronic band magnetism, such as Pd. Magnetic susceptibility measurements provide no evidence for localized magnetic moments. That is, neither the temperature nor magnetic-field dependences of measured susceptibilities provide evidence for disordered or ordered moments.

**B**    **Specific-heat measurements**

Studies of the specific heat of plutonium were initially (before about 1975) concentrated on the determination of the structural phase transitions that are known to take place between room temperature and the melting point. Dean and coworkers made one of the first measurements above room temperature of the α-phase.[44,45] The data in the latter reference were republished by Kay and Loasby[46] in a more comprehensive paper that at present stands as the highest quality data above room temperature. Specific-heat measurements and the enthalpy curve for the δ-phase (1 wt. % Ga) were reported by Rose and coworkers using an ice calorimeter.[47]

Not surprisingly the first calorimetric measurements of Pu below 300 K were made under the cloak of the Manhattan Project to determine the $^{239}$Pu half-life and were subsequently published in 1947 by Stout and Jones.[48] Nearly a decade later more measurements on plutonium below 300 K were published in an effort to determine the Sommerfeld coefficient ($\gamma$) and low-temperature limiting Debye temperature ($\Theta_D$). Unusual effects were observed in the α-phase[27] and were attributed to antiferromagnetic ordering. However, the peaks were shown to be measurement artifacts (adsorbed helium on the sample surface) by Taylor et al.[30] In another study Taylor and Linford[29] showed more evidence ruling out the existence of a magnetic ordering transition of the type found in α-Mn, Ref. 41.

Recent measurements,[49] shown for δ-Pu in **Fig. 4**, report on well-characterized samples of both α and δ-Pu. A key factor illustrated by this study, and observed directly by optical metallography, neutron diffraction, and elastic constant measurements[46] made on the same sample, is that at low temperatures there is a formation of a monoclinic martensite phase (called α') in the stabilized (in this case by 5% Al) δ-Pu phase. Thus, the anomalies in the specific heat are now attributed almost exclusively to *structural* effects and not to ordered magnetism, in δ-Pu. This study also made use of recent phonon density of states measurements[50] with neutron inelastic scattering data to subtract accurately the phonon contribution.

A major problem in trying to extract Sommerfeld values is the self-heating, which is nominally 2 mW g$^{-1}$ for the $^{239}$Pu nucleus. In practice it is extremely difficult to achieve temperatures much below ~ 2 K unless very small (< mg) samples are used, and even then there is always some doubt as to the real temperature of the sample. Nevertheless in a recent calculation, Harrison obtains remarkable agreement the γ values for the light actinides using a modified solution to a two-electron problem.[51] The experimental values are summarized and are shown here in **Fig. 5**. The values for α- and δ-phases are in favorable agreement with previous values determined by Gordon *et al.*[30] for α-Pu, and by Stewart and Elliott[52] for different alloy concentrations of δ-Pu. The values of the Sommerfeld coefficient for the light actinides were obtained from the following references. Thorium measurements made by Griffel and Skochdopole,[53] Pa measurements by Stewart *et al.*[54], low-temperature U measurements by Bader *et al.*[55], Np measurements by Gordon *et al.*[30], α- and δ-Pu by Lashley *et al.*[49], α'-Pu by Stewart and Elliott[52], and Am measurements by Mueller *et al.*[56].

A further indication of whether any excess entropy associated with magnetism is involved in these phase transitions would be the sensitivity of the specific heat to applied magnetic field.[57] That is to say that it is conceivable (in light of the theories promoting antiferromagnetic fluctuations) that variations in γ with magnetic field could occur. For α-Pu the results are shown in **Fig. 6** and for δ-Pu (2 wt % Ga) recent measurements made by a thermal relaxation technique described elsewhere[58] are shown in **Fig. 7a.** No differences have been found in either phase. At low-temperatures if α-Pu were located near a magnetic boundary one might expect γ to show a magnetic field dependence; however, the γ at zero field was found to be 17(1) mJ K$^{-2}$ mol$^{-1}$ and at 14 T was found to be 16(1) mJ K$^{-2}$ mol$^{-1}$. Similarly, γ obtained at H = 0 T and H = 9 T for δ-Pu (2 wt % Ga) are found to be within experimental error of one another as shown in Fig. 7b. These results alone do not prove completely that plutonium is non magnetic. If the critical temperature is 42 K an estimate of the field, based upon cyclotron resonance required to couple to the anomaly and assuming a *g*-factor = 2, would be of the order of 60 T. However, they provide no support for ordered magnetism or even of *any* magnetic entropy in the system, as would be associated with either static or dynamic disordered moments. It is the only specific-heat measurement of plutonium in a magnetic field over a large temperature range.

In summary, the specific-heat measurements indicate the absence of magnetic entropy. The Sommerfeld coefficient γ does not couple to magnetic field in either the α- or δ-phases of Pu up to magnetic fields of 14 and 9 T in the α– and δ-phases, respectively. The peak in *C/T* in the δ-Pu samples below ~ 50 K (see Fig. 4 inset) has been identified as a martensitic transition rather than antiferromagnetic ordering as suggested previously. As for the Schottky effect it is possible that it could arise from structural defects, caused by self-irradiation damage, similar to but from different origins to the structural-defect Schottky known to exist at low temperatures in copper.[59]

**C Neutron-elastic scattering**

Given the unusual thermal properties of the different phases in Pu, one of the important questions is the role of anharmonic forces in this unusual material. Starting about a decade ago, experiments were conducted on plutonium in various forms at both IPNS (at Argonne National Laboratory) and the LANSCE (at Los Alamos) neutron sources to measure the diffraction patterns from polycrystalline samples as a function of temperature from 5 to 800 K. A recent account of the results of this work is given by Lawson *et al.*[60] In the present paper we shall not be concerned with the results of this study per se, interesting though they are, but we note that this extensive data set of neutron diffractograms can also be used to search for both ordered antiferromagnetism (i. e. new peaks in the diffractograms) or diffuse scattering from disordered moments, which would appear in the background of the patterns. Since these experiments were performed in a "diffraction" mode with no analyzer, they cannot distinguish between static or dynamic disordered moments.

Before showing a series of such diffractograms and comparing them with various theoretical predictions for the ordered antiferromagnetism in Pu, we need to discuss the form factor expected for the dipole moment of Pu. The neutron is sensitive to the dipole moment at the atomic site through its interaction with the dipole moment on the neutron, and this interaction, besides being a vector, also depends on the spatial extent of the magnetic moment centered at the atomic site. Normally, these form factors, abbreviated as f(Q), where Q = |**Q**| is the momentum transfer [Q = 4π(sinθ)/λ = 2π/d, where θ is the Bragg angle, d is the d-spacing of the atomic planes, and λ is the wavelength of the scattered radiation] of the scattering process, have a maximum of unity at Q = 0, and fall away in a regular manner as Q increases in value. (Since we are using the dipole approximation[61] the directional aspect of the momentum transfer is not considered.) However, in the case that the orbital and spin moments are oppositely opposed the situation is more complex. In this case, as discussed at length by Lander,[62]

$$\mu f(Q) = \mu (\langle j_0 \rangle + C_2 \langle j_2 \rangle + \ldots) \qquad (1)$$

where $\mu$ is the *total* moment, $\langle j_i \rangle$ are Bessel functions derived from the single-electron probability function (in this case in the 5$f$ shell), and $C_2$ is a constant given by

$$C_2 = \mu_L/\mu \qquad (2)$$

where $\mu = \mu_L + \mu_S$ with the latter being the respective orbital and spin moments. The $\langle j_0 \rangle$ function has a value of unity at $Q = 0$ and falls slowly with increasing Q, whereas the $\langle j_2 \rangle$ function has a value of zero at $Q = 0$ and increases to a maximum value of ~ 0.2 at $Q \sim 4.5$ Å$^{-1}$.

If we now consider the case of Russell-Saunders coupling (which is valid for the rare earths) then $\mu_S = 2(g-1)J$ and $\mu_L = (2-g)J$ where $g$ is the Landé splitting factor and J is the total angular quantum number. For Sm$^{3+}$ (the 4$f^5$ analog of Pu$^{3+}$) with $g = 2/7$ and $J = 5/2$, $\mu_S = -25/7$, $\mu_L = +30/7$, $\mu = +5/7$, $C_2 = 6.0$. This is clearly an extraordinary f(Q) as its must have a large maximum near the maximum of $\langle j_2 \rangle$ and has indeed been observed by Koehler and Moon[63] for Sm metal. More details of other Sm compounds are given in Ref. 62.

Such a simple analysis will not be totally relevant for the actinides; we know for a start that intermediate coupling must be present and this increases the $g$ factor for Pu$^{3+}$ from 0.287 to 0.414. In Eq. (1) and (2) the resulting $C_2$ is 3.8. Such a value has been observed[64] in a *localized* Pu compound PuSb, which exhibits a total moment $\mu \sim 0.7$ $\mu_B$. The large $C_2$ giving a characteristic *hump* in f(Q) may be visualized another way as f(Q) is the Fourier transform of the magnetization. The spatial dependence of the orbital and spin magnetizations are different around the nucleus (the orbital is actually more contracted in real space) and when these are subtracted this gives a doughnut effect. If $|\mu_S| = |\mu_L|$ and they are oppositely directed, then the total moment $\mu = 0$, but the difference in their spatial extent *would still allow a measurable signal to be seen in neutron scattering*. This is a crucial point in discussing theories that have emphasized the cancellation of the spin and orbital moments. A signal is still seen in neutron scattering. A good example is the uranium moment in UFe$_2$. In this material the total moment on the U atom is ~ 0.01 $\mu_B$, but the individual $\mu_S$ and $\mu_L$ are almost equal and opposite and about 0.22 $\mu_B$. The neutron experiments[65] observe a maximum of ~ 0.05 $\mu_B$, some 5 times larger than the total moment because of the differing spatial extents of the two distributions as discussed above.

There is, however, a further point of the theory that we need to emphasize again. In all Russell-Saunders coupling schemes for a less than half-filled shell $|\mu_L| > |\mu_S|$, but the theory we have discussed in Sec. II repeatedly concludes with the opposite, i.e $|\mu_L| < |\mu_S|$. In this situation, taking as a representative example, for the prediction of $\mu_S = 4.5$ $\mu_B$ and $\mu_L = -2.5$ $\mu_B$, giving $\mu = 2.0$ $\mu_B$ we find $C_2 = -1.25$, and a completely different f(Q). It should be stressed that in all the experiments so far on U, Np, and Pu systems *no such unusual* f(Q) has ever been found.[62,66] These predictions are incorrect as they do not take into account orbital polarization effects; when these are considered we find agreement with the experimentally observed fact that $|\mu_L| > |\mu_S|$.

In the simulations below, therefore, we have used a f(Q) derived with $C_2 \sim 4$. Experiments with strongly hybridized systems, for example $PuFe_2$,[67] have shown that the $C_2$ can reach $\sim 6$, so that a factor of 4 is a good compromise for $Pu^{3+}$.

Neutron-diffraction data have been obtained on various plutonium samples prepared from material enriched to 95% $^{242}Pu$. We have used these data to test some of the magnetic models given in the literature. **Figures 8 (A-D)** show Rietveld refinements of neutron-diffraction data from various phases of plutonium. Fig. 8A shows the data for δ-phase Pu stabilized to low temperature with 5 at. % Al. The points are the observed diffraction data taken on the HIPD instrument at the Manuel Lujan Jr. Neutron Scattering Center at LANSCE at the Los Alamos National Laboratory. Magnetic Rietveld refinements were done using the formalism based on Shubnikov magnetic space groups in the GSAS program.[68] The points are the diffraction data observed at 15 K normalized by the incident spectrum. The line through the data is a pattern calculated from the model specified by Söderlind[15] with Shubnikov group B2cm and an ordered moment of 1 $\mu_B$. The line immediately below the diffraction pattern (magenta on-line) is the difference between the refined Söderlind model, which gives $\mu_B < 0.02 \pm 0.40$ $\mu_B$, and the observed data. The line further down (blue on-line) is the difference between the Söderlind model with the moment fixed at 1 $\mu_B$ and the observed data. Next is a row of arrows showing the calculated positions of the magnetic reflections. Finally, at the bottom, is a row of arrows showing the calculated positions of the nuclear reflections. The dotted line shows the value of the $Pu^{3+}$ magnetic form factor.

Figures 8(B-D) are organized in a similar way, with differences explained in **Table I**. This table gives the limits of ordered magnetism that can be extracted from the fits; they are all below the level of $\sim 0.4$ $\mu_B$, which is generally regarded as the limit that can be excluded when doing neutron diffraction on polycrystalline samples. The high-temperature data were obtained at temperatures very much higher than the Debye-Waller temperatures[69] of 80 and 71 K, for the δ- and ε-phases, respectively; it was therefore necessary to fit the temperature diffuse scattering, and this was done using a scheme described in the literature.[70]

No excess background that could be associated with paramagnetic scattering from disordered local moments (either static or dynamic) was found at large d-spacings (small Q) for any of the phases.

**D Neutron-inelastic scattering**

Neutron-inelastic scattering has the capability to measure time-dependent fluctuations of magnetic moments. With the absence of any spatial correlations in the distribution of moments, the inelastic scattering spectrum can be derived from the single-moment fluctuations of a paramagnet. This cross-section is wellknown and obeys the following relationship

$$S(Q,\omega) = \frac{1}{2\pi}\left(\frac{\gamma r_0}{\mu_B}\right)^2 f^2(Q)\left[1 - e^{-\beta\omega}\right]^{-1}\chi''(\omega)$$

where the symbols have their usual meanings.[61,62] The imaginary part of the local magnetic susceptibility has a quasielastic Lorentzian response proportional to the squared moment.

$$\chi''(\omega) = \frac{1}{3}g^2\mu_B^2 J(J+1)\left(\frac{\hbar\omega}{kT}\right)\frac{\Gamma/2}{(\Gamma/2)^2 + (\hbar\omega)^2}$$

In principle, the neutron-diffraction results discussed in the previous section are sensitive also to dynamically fluctuating moments, since diffraction measures the energy integral of $S(Q,\omega)$. However, neutron inelastic scattering data can unambiguously determine the quasielastic spectrum and the characteristic fluctuation energy, $\Gamma/2$. In properly normalized inelastic spectra, the size of the disordered moment can be determined as well.

We recently performed neutron inelastic scattering measurements[50] on a polycrystalline sample of δ-phase $^{242}$Pu$_{0.95}$Al$_{0.05}$ using the PHAROS spectrometer at the Lujan Center at Los Alamos National Laboratory. (Details of the experimental setup are given by McQueeney *et al.*[50]) This measurement was optimized for phonon studies at large Q (large scattering angles), but the PHAROS instrument measures scattering angles from 2-145 degrees simultaneously, so we searched in the low-Q (small scattering angle) region for signs of magnetic quasielastic scattering indicative of dynamicallydisordered moments. The left-hand panel of **Fig. 9** shows data summed over scattering angles from 10-30 degrees, corresponding to $0.75 < Q < 2$ Å$^{-1}$, for various temperatures. A temperature-dependent signal is observed that corresponds exactly to the expected weak low-Q phonon cross-sections calculated from a lattice dynamical model for δ-Pu and shown in right-hand panel of Fig. 9. Also shown in the right-hand panel of Fig. 9 are the calculated magnetic quasielastic scattering cross section at various temperatures for $g^2 J(J+1) = 0.5$ and $\Gamma/2 = 5$ meV (estimated from the Sommerfeld constant of δ-Pu). The comparison shows that the low-angle intensity is mainly, if not completely, explained by phonons. Even though the magnetic form factor of Pu, as discussed above, may be unusual, the form factors shown in Fig. 8 have non-zero values in the range of these experiments..

*The combination of neutron elastic (Fig. 8) and inelastic (Fig. 9) scattering data shows no convincing evidence for either long-range ordered or disordered (static or dynamic) magnetic moments.*

**IV    DISCUSSION**

In this paper we have demonstrated beyond all reasonable doubt that there is no *ordered* magnetism involving the 5*f* electrons in Pu metal in any of its crystallographic phases and down to a base temperature of ~ 4 K. The experimental evidence presented includes magnetic susceptibility, specific heat (with an applied field of up to 14 T), NMR, and neutron scattering, both elastic and inelastic. A recent NMR experiment using the Ga signal on 1.5 wt % Ga stabilized δ-Pu reported by Piskunov et al.[71] confirms the earlier work[32] that there is no evidence for ordered magnetism from NMR. Almost simultaneously with the above paper, an NMR study from Los Alamos National Laboratory by Curro and Morales[72] has further excluded the possibility of ordered magnetism in δ-Pu.

Previous reports of "anomalies", seen especially in the specific heat, are now ascribed to structural effects, most probably the occurrence of martensitic transformations of some parts of the sample(s). The lessons of α-U, such as dealing with temperature hysteresis and self-heating below 1 K are salutary in this respect. For many years magnetism was suspected in this material at the famous 43 K transition, but it was finally established[73] some 25 years ago that α-U undergoes a series of charge-density wave (CDW) transitions, and that there is no sign of any magnetism, either ordered or disordered. Whether such a CDW occurs in either α- or stabilized δ-Pu is beyond the scope of the experiments described here. It is for example, difficult to see the CDW in α-U with diffraction experiments on polycrystalline samples, and, in our opinion, the questions on a possible CDW in Pu will be answered only by either electron or synchrotron x-ray diffraction at low temperatures. The recent elegant experiments[74] on the phonons in δ-Pu by inelastic x-ray scattering open a new perspective for such a discovery.

A more difficult question is whether *disordered* magnetic moments could exist and perhaps order at a much lower temperature, which is experimentally inaccessible in Pu given the self-heating of most Pu isotopes. The most compelling evidence against such a scenario is the susceptibility and the neutron inelastic scattering results. In contrast to the arguments advanced by Méot-Raymond and Fournier[43] about an "effective" moment showing disordered local moments, we have argued that the *overall* shape of the susceptibility, as well as its large value in all phases, (Fig. 3), gives no support to the legitimacy of such an argument. Such small temperature-dependent effects can easily be derived from a system with wide 5$f$ bands, without recourse to disordered local moments.

The absence of any diffuse scattering in the neutron-diffraction patterns (except at high-temperature from thermal disorder), Fig. 8, also argues against any disordered local moments. These measurements integrate over the whole incident neutron spectrum, in this case up to perhaps as much as 100 meV, depending on the instrumental conditions. Knowing the γ coefficients of α- and δ-Pu (17 and 65 mJ K$^{-2}$ mol$^{-1}$, respectively, see Fig. 5), we can roughly deduce the spectral width of the quasielastic scattering that would be associated with such dynamic (i.e. fluctuating) moments involving 5$f$ states.[75] Since the half-width at half maximum (HWHM) of the neutron scattering response (Γ/2) is roughly proportional to 1/γ the spectral response (Γ/2) would be expected to be 20 meV in the case of α-Pu, and 5 meV in the case of δ-Pu. Even if a parameter such as (Γ/2) is not related directly to γ for 5$f$ systems,[75] we know that in this energy range paramagnetic scattering should be seen from fluctuating moments. Such energies are *exactly* the range of the neutron inelastic scattering experiment already performed,[50] and the resulting scattering around the elastic scattering position is shown in Fig. 9. Again, there is no evidence for such a magnetic signal. A new experiment optimizing conditions at low-scattering angle and being able to place the scattering on an absolute scale with respect to Bohr magnetons is planned.

In each of the experiments reported one can, of course, find reasons that a signal from disordered moments might have been missed, but the failure of *all* techniques in this respect leads to the Occam's razor conclusion that such disordered moments do not exist. This should not, of course, be taken to imply that conduction electrons involving presumably 6*d*–7*s* and maybe even 5*f* states are absent; Pu is in all senses a metal.

In reviewing much of the published theory on plutonium over the last 15 years[4-24] it is clear that the major focus has been the *volume* difference between the α- and δ-phases. To a large extent this problem has been solved satisfactorily by modern calculations. Many other properties, such as the elastic and thermal effects have also been explained by these efforts. Unfortunately, however, the predictions in many (but not all) calculations of relatively large localized moments, especially in δ-Pu, are *not* in agreement with experiment.

Furthermore, the situation in δ-Pu may be more interesting than many theories predict. The real question is to understand *why* the 5*f* electrons do *not* order magnetically, or *why* such ordering cannot be observed by all the techniques that have been successful to detect magnetism in the last half century. We return to two theories that may be relevant in this context. First, Nordström and Singh[9] proposed an intra-atomic noncollinear magnetization applicable to plutonium. Although we believe that such a magnetization would be observable by neutron scattering, for example as discussed by Blume[76] in 1963, it is possible that single crystals might be needed and rather original experiments. More work to understand whether an experiment is justified, and what exactly that should be (neutrons or resonant x-ray scattering) is required. Second, the results obtained by the dynamical mean field theory of Savrasov *et al*.[15] imply, at least to us, that the local moments are "washed" out over short time scales and thus may not be observable to probes such as NMR and neutron inelastic scattering, depending on the probes' observational frequency window. Again, more effort to quantify these predictions for experiments on Pu would seem worthwhile. Towards this end specific-heat measurements with magnetic fields up to 45 T at the National High Magnetic Field Laboratory are also underway.

**Acknowledgments**

We thank Merwyn Brodsky, Ladia Havela, Sig Hecker, Gabriel Kotliar, Bill Nellis, Peter Oppeneer, and Jim Smith for a number of conversations or correspondence about this subject. GHL thanks the staff at MST Division and LANL Laboratory Management for the award of a John Wheatley Scholarship, during which period at LANL much of the research for this article was developed.

Part of this research is sponsored by the US–Department of Energy (DOE) National Nuclear Security Administration and the Office of Science. The work has benefited from the use of the Intense Pulsed Neutron Source at Argonne National Laboratory and of the Los Alamos Neutron Science Center at Los Alamos National Laboratory. IPNS is funded by the US–DOE, BES-Materials Science, under Contract W-31-109-ENG-38, LANSCE is funded by US–DOE under Contract W-7405-ENG-36, and

**Table I**

Results of refinements as shown in Fig. 8. In each case a model associated with a theoretical prediction for magnetic ordering is the starting point for the Rietveld analysis and the final moments obtained are shown in the last column.

| Figure | phase | T (K) | Instrument/ Source | Schubnikov group | Ref. | initial moment ($\mu_B$) | refined moment ($\mu_B$) |
|---|---|---|---|---|---|---|---|
| A | δ-Pu 5 at. % Al | 15 | HIPD/ LANSCE | B2cm (Söderlind) | 18 | 1 | 0.02 ± 0.4 |
| B | δ-Pu | 713 | GPPD/ IPNS | Simple AF I´4/mm´m´ | 17,21 | 2 | 0.04 ± 0.09 |
| C | ε-Pu | 768 | GPPD/ IPNS | Simple AF I´4/mm´m´ | 17 | 2 | 0.1 ± 0.4 |
| D | α-Pu | 15 | HIPD/ LANSCE | AF P2´$_1$/m | 21 | As in Ref. 21 | — |

**Figure Captions**

**Figure 1 (color on-line)**
Atomic volumes of the 3$d$, 4$f$, and 5$f$ elements as a function of the electron count, increasing to the right. Note the parabolic shape of the 3$d$ series, the almost constant values of the rare-earths (Eu and Yb are divalent as metals and have a much larger volume than the other trivalent metals), and the unusual behavior of the 5$f$ elements, with a minimum volume near Pu, and a very large increase between Pu and Am.

**Figure 2**
Thermal expansion in the different phases in plutonium as a function of temperature. Six phases exist in plutonium, with the change from α- to β- having the largest discontinuity in volume. The crystal structures (number of unique atom types) are α-Pu monoclinic (8); β-Pu monoclinic (7); γ-Pu orthorhombic (1); δ-Pu fcc (1); δ'-Pu body-centered tetragonal (1); ε-Pu bcc(1).
The more conventional behavior of an element like iron is shown for comparison as a broken line. Note that this figure appears in Ref. 1.

**Figure 3 (color on-line)**
Molar susceptibility of Mn (taken from Ref. 41) and various forms of Pu plotted as a function of the melting point (1519 K for Mn and 913 K for Pu), see text for details of Pu data.

**Figure 4 (color on-line)**
The separation of the experimental specific heat of δ-Pu into its phonon and electronic components based on a knowledge of the phonon density of states derived from Ref. 50. The inset shows the evolution of $C_{el}/T$ with temperature; notice how this returns to 17 (the low-temperature α value) by 300 K. (Adapted from Ref. 49)

**Figure 5 (color on-line)**
Evolution of the Sommerfeld coefficient (electronic specific heat) for the light actinides showing an electronic transition in the series. The spline curve (dashed) indicates the γ values for the α-phase of each actinide element. The measured values shown on the curve are in agreement with those calculated in Ref. 51.

**Figure 6 (color on-line)**
The specific heat of α-$^{239}$Pu in zero magnetic field and 14 T from 15 to 120 K to illustrate the large bump centered at 40 K that has been sometimes referred to in the literature as a Schottky effect.

**Figure 7 (color on-line)**
(a) The specific heat of δ-$^{239}$Pu (2 wt % Ga) in zero magnetic field and H = 9 T from to 150K. The entropy, obtained from integration of C/*T* with respect to *T* is shown on the right axis.
(b) The low-temperature limiting Debye temperature ($\Theta_D$) and Sommerfeld coefficient (γ) are shown for zero magnetic field and H = 9 T. One can see that the difference between these curves are within the experimental error (shown as error bars) and therefore shows no coupling of the specific heat to field at least to 9 T at 4 K.

**Figure 8 (color on-line)**
Neutron-diffraction data and analyses for various phases of Pu at different temperatures. Details are given in the text and in Table I. The dashed line is the theoretical magnetic form factor for Pu$^{3+}$ as discussed in the text.

(A) δ-Pu (5 at. % Al) at 15 K.
(B) Unalloyed δ-Pu at 713K.
(C) Unalloyed ε-Pu at 768K.
(D) Unalloyed α-Pu at 15K.

**Figure 9 (color on-line)**
The left-hand panel shows neutron inelastic scattering data that has been averaged over the low-angle range from 10-30 degrees. For details of experimental conditions, see Ref. 50. The right-hand panel shows in absolute units the expected phonon scattering (solid lines) in this angle/temperature range as calculated from a simple lattice dynamical model for δ-Pu. The magnetic quasielastic scattering from a paramagnet (dotted lines) are also shown in the right-hand panel; parameters for the model are given in the text.

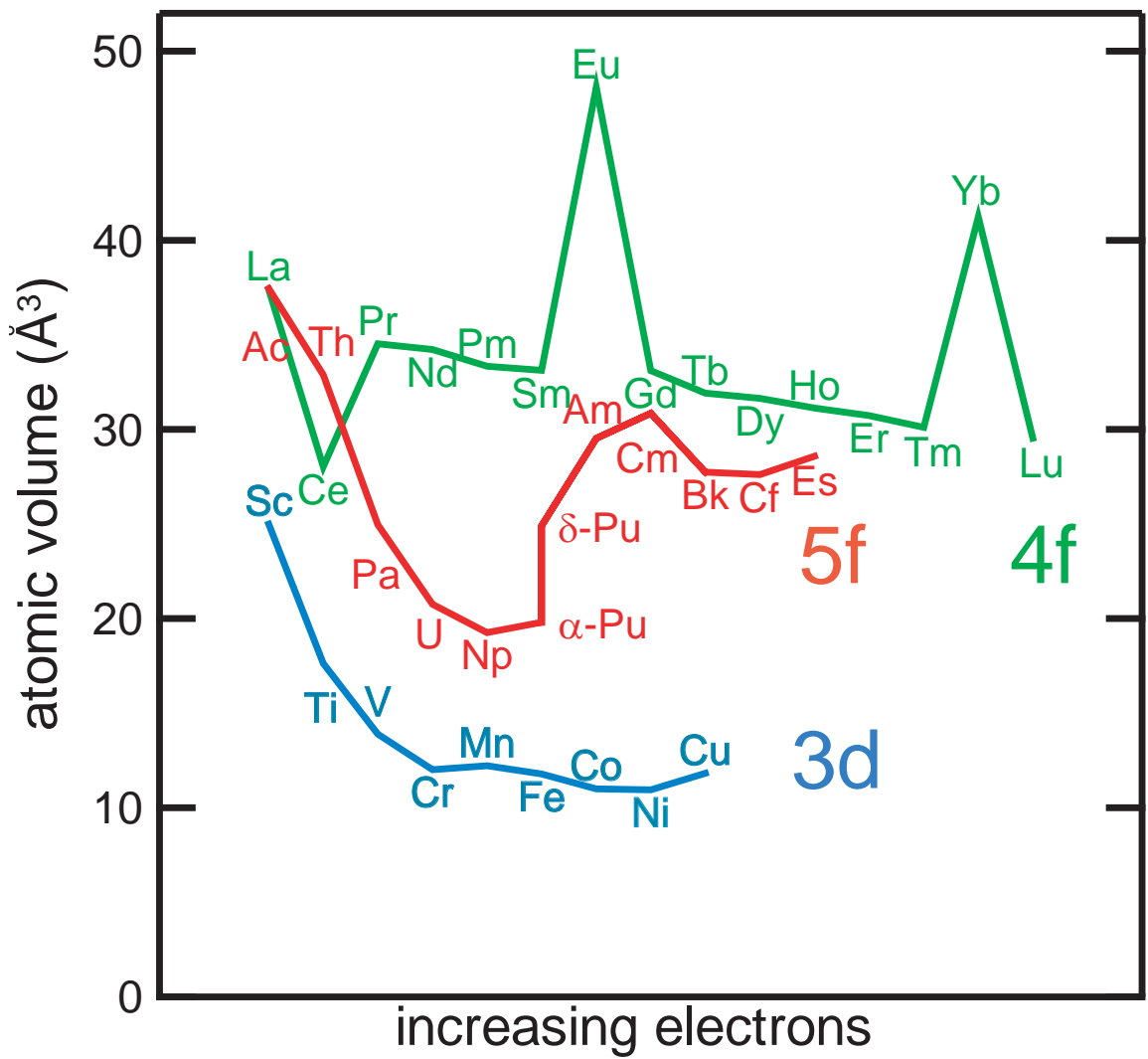

Figure 1

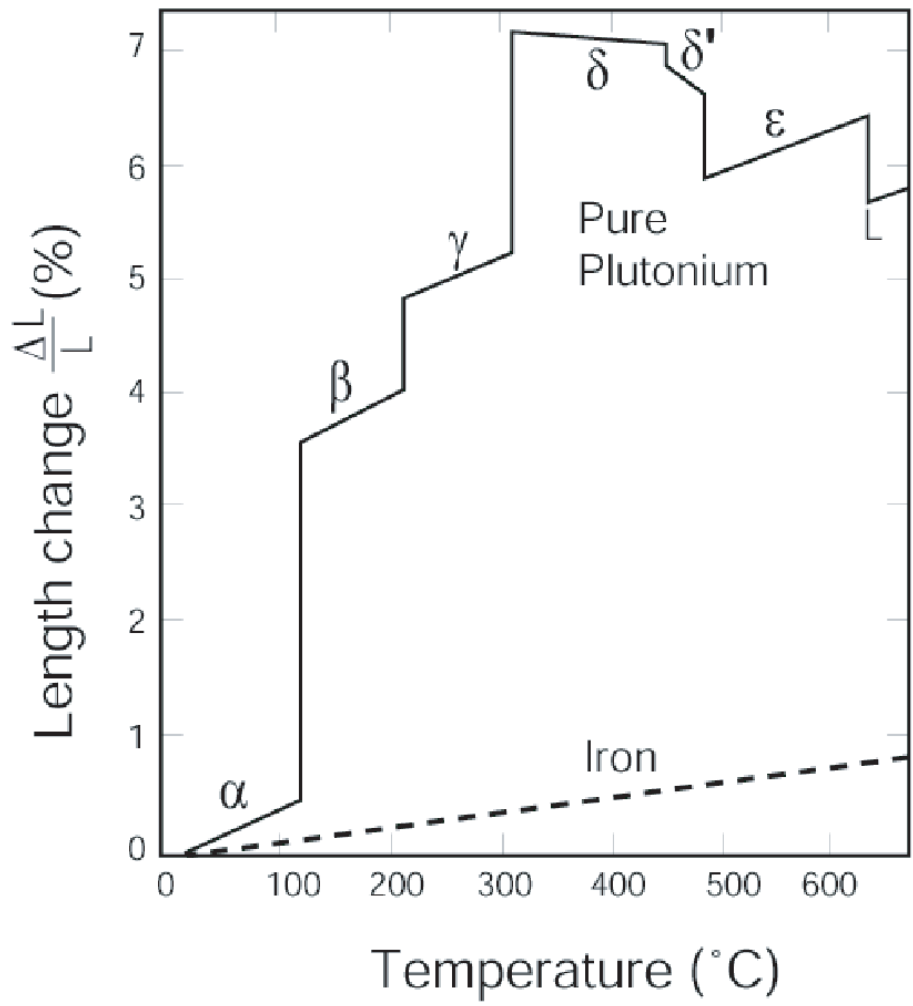

Figure 2

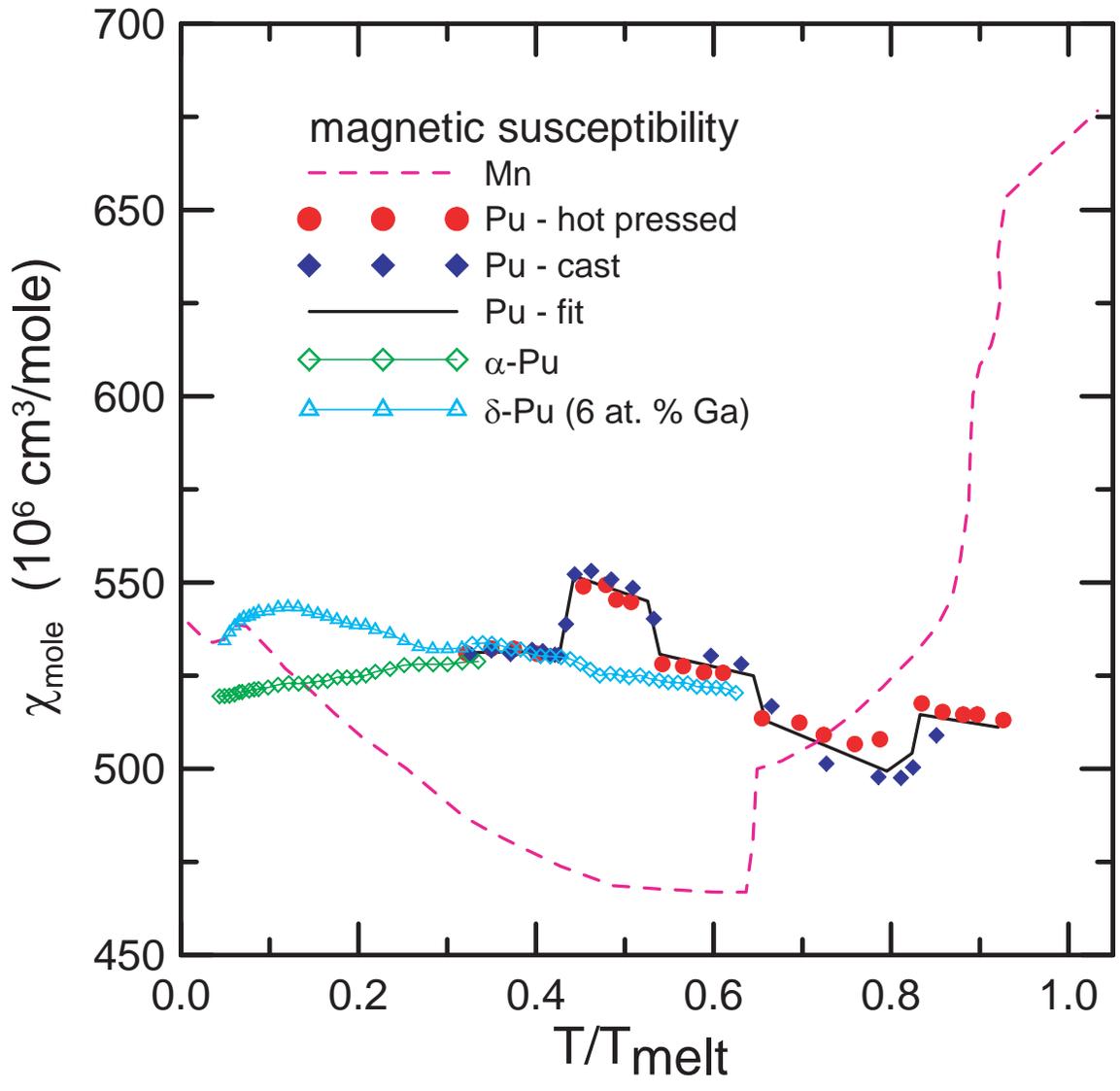

Figure 3

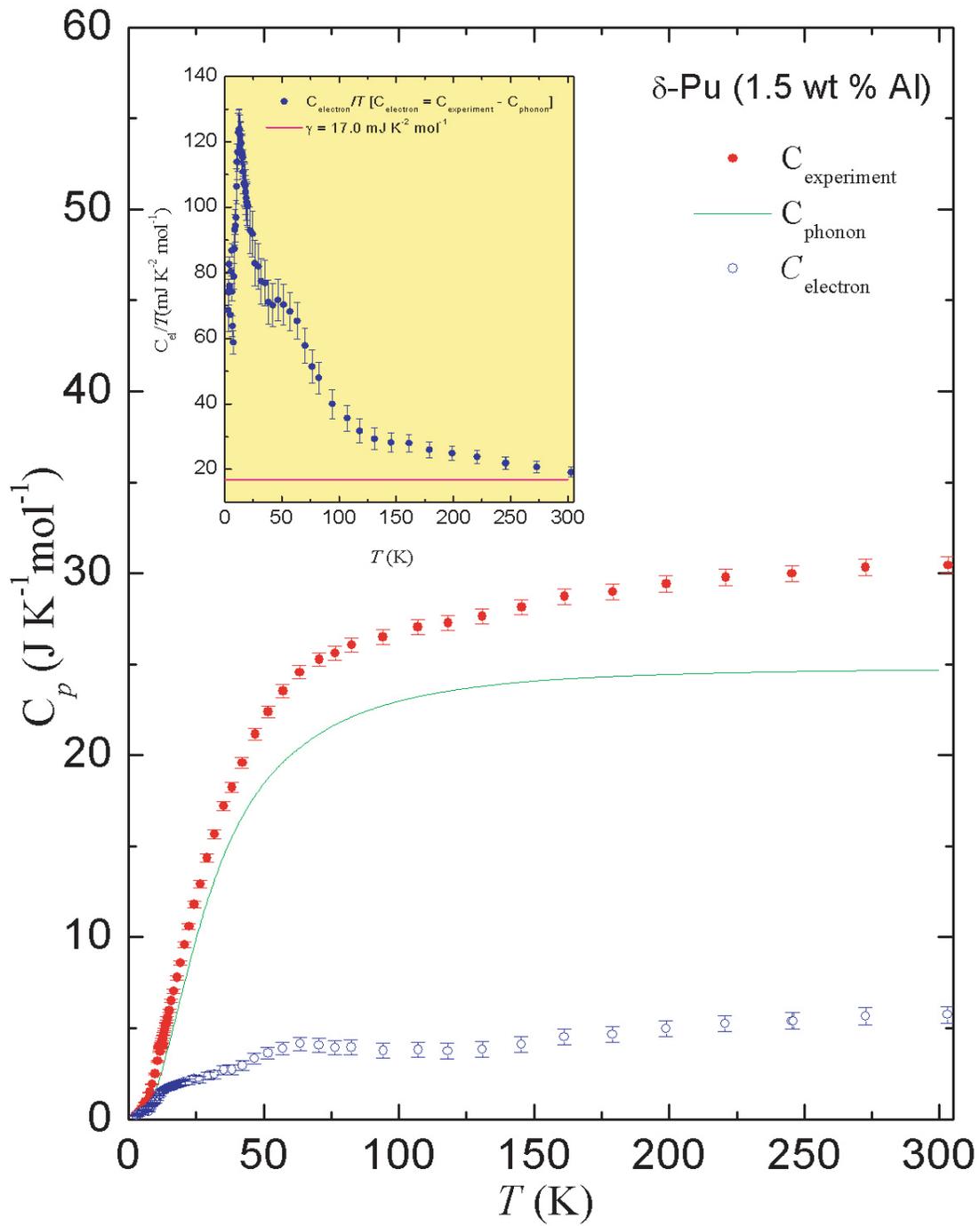

Figure 4

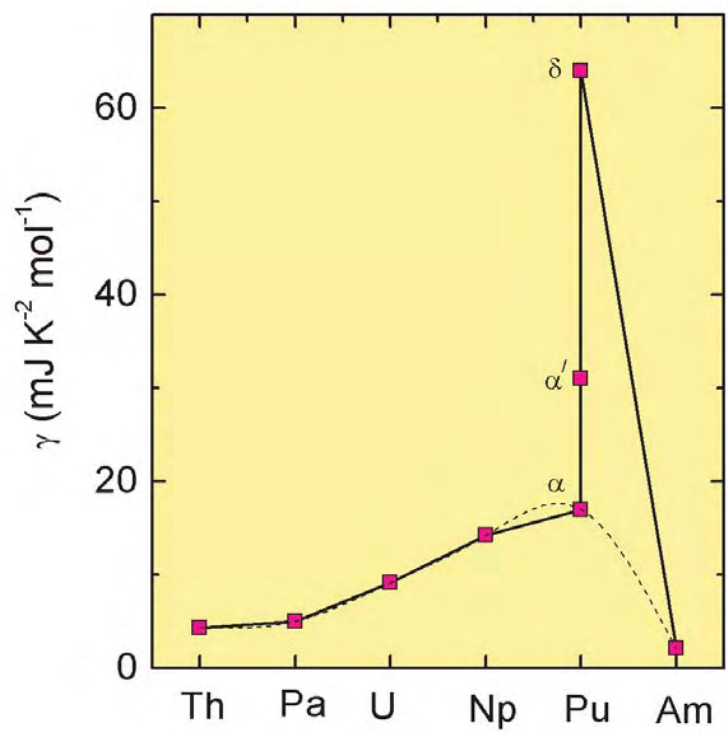

Figure 5

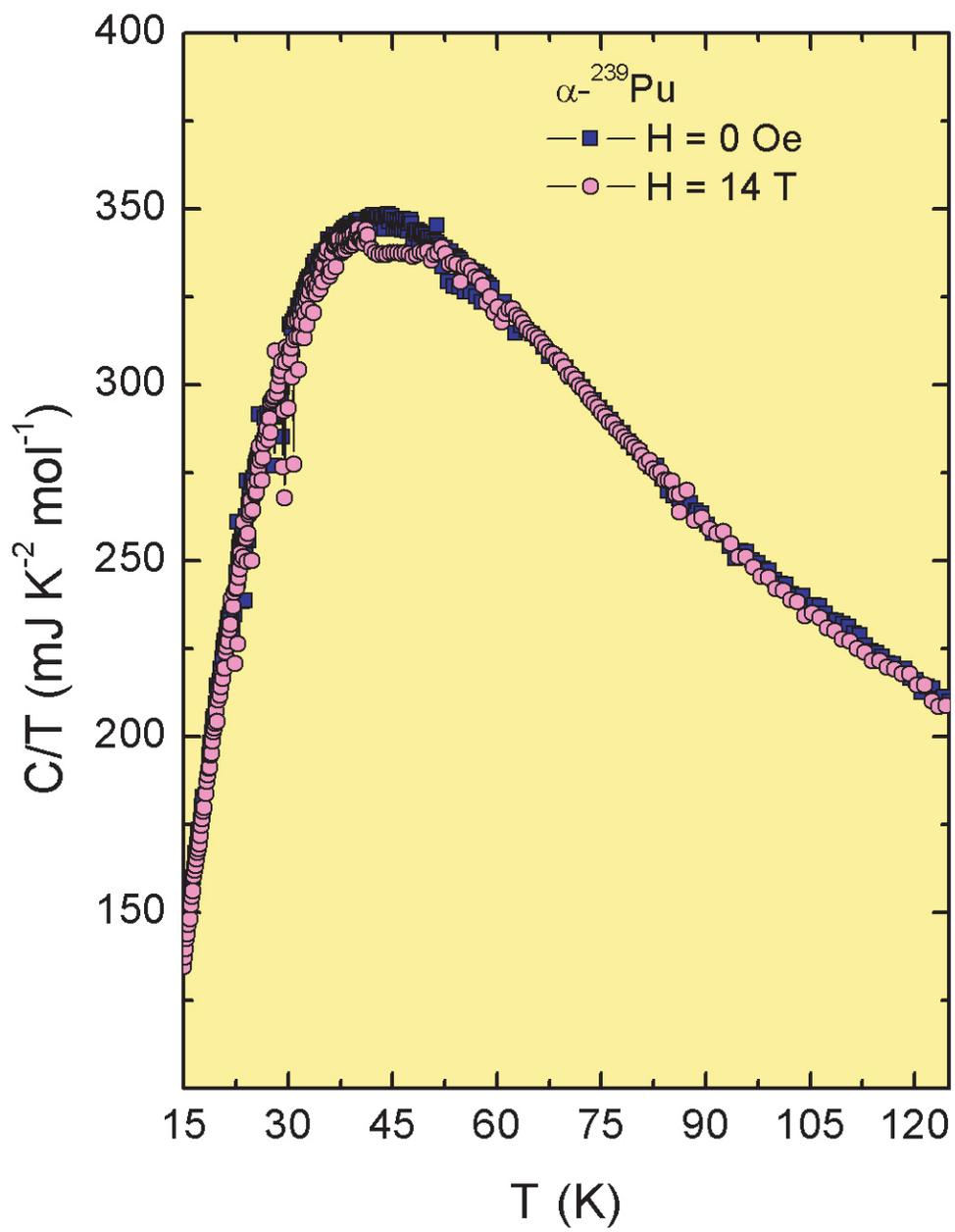

Figure 6

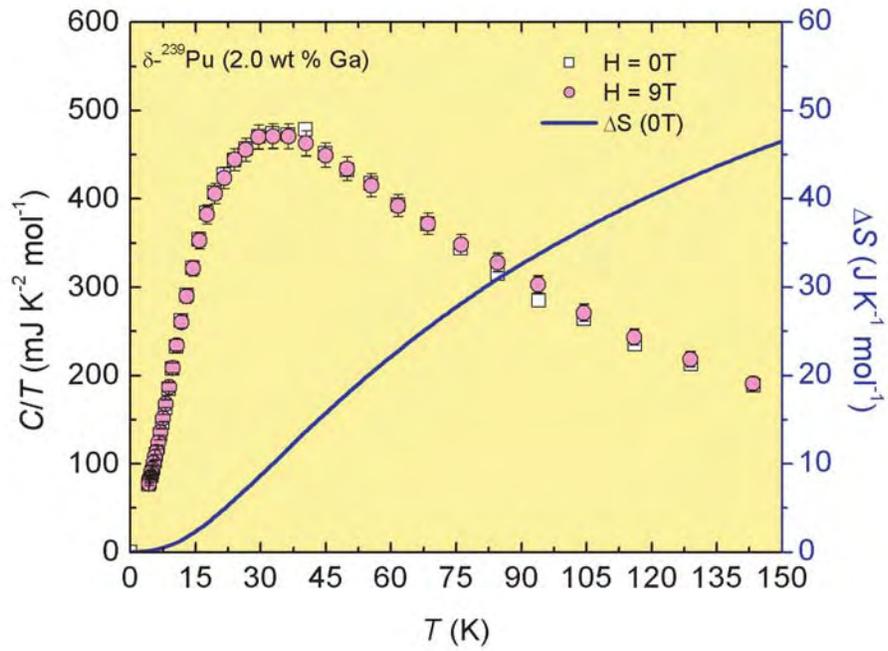

Figure 7a

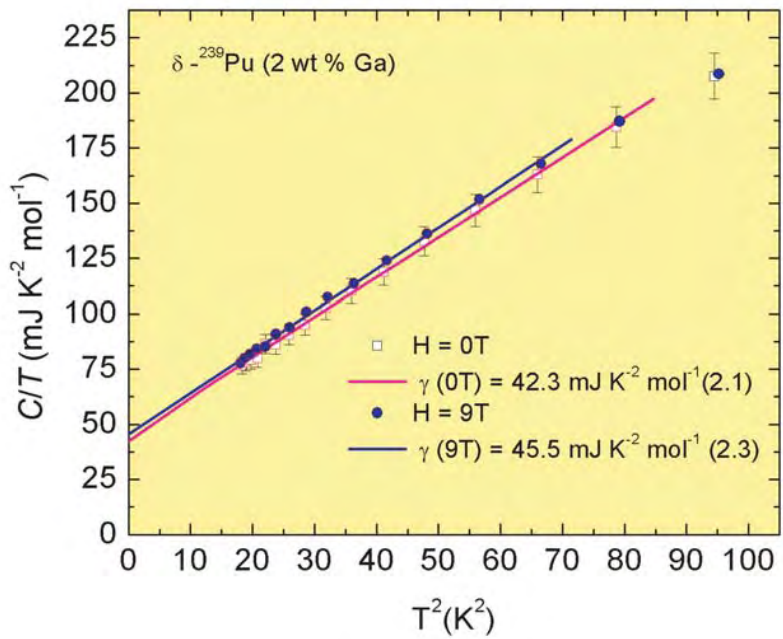

Figure 7b

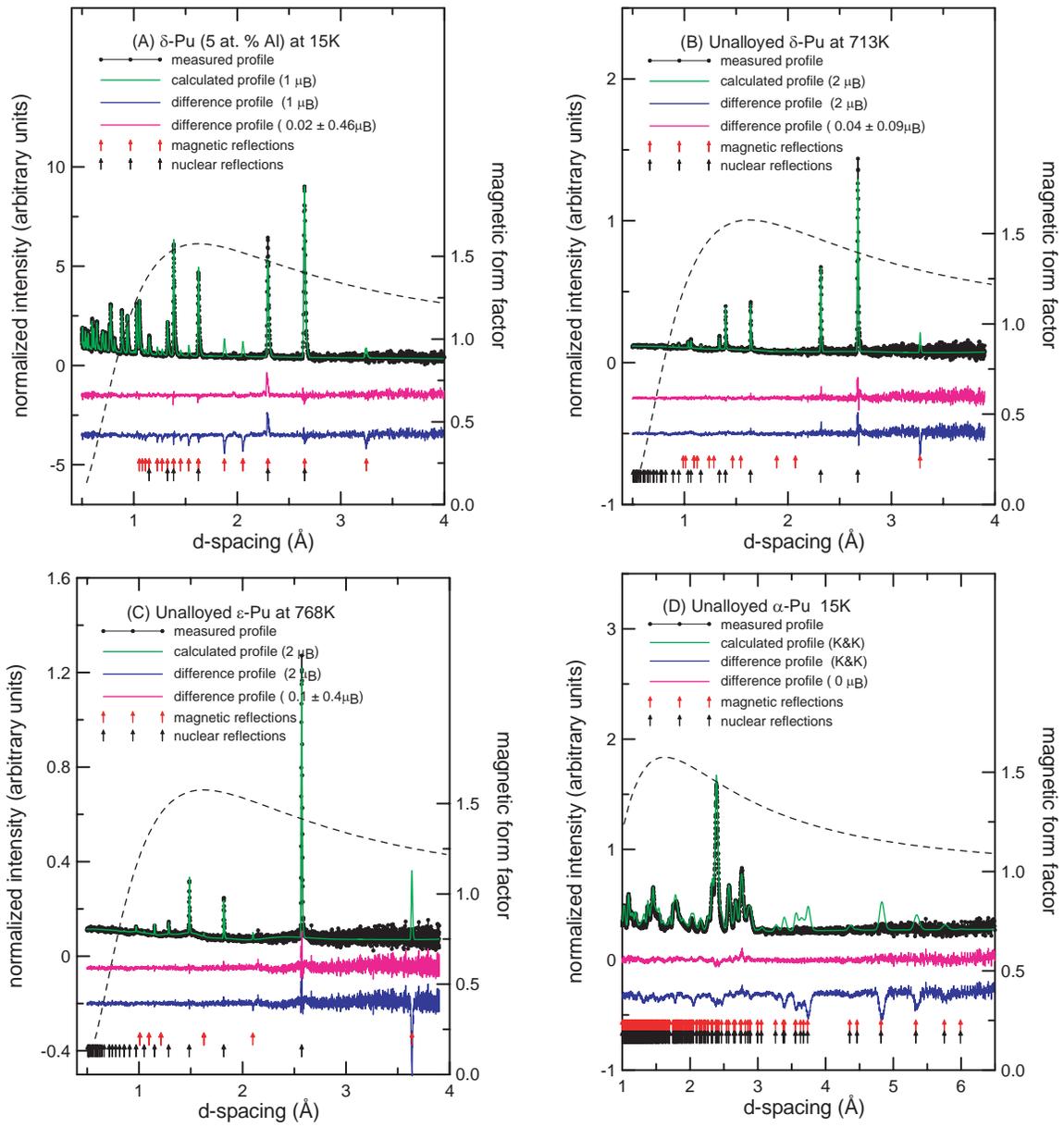

Figure 8

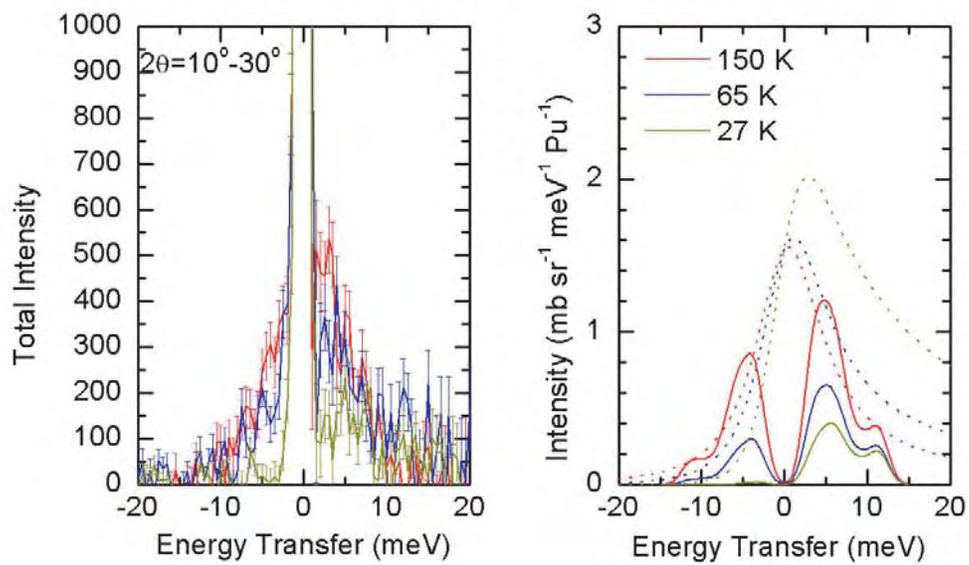

Figure 9